\documentclass[authoryear,preprint]{elsarticle}
\usepackage{graphicx}
\usepackage{subfigure}
\usepackage{natbib}

\journal{New Astronomy}
\begin{document}

\begin{frontmatter}

\title{Gravitational Microlensing: A parallel, large-data implementation$^1$}

\author[]{Hugh Garsden}
\ead{hgarsden@physics.usyd.edu.au}
\author[]{Geraint F. Lewis}
\ead{gfl@physics.usyd.edu.au}

\address{Sydney Institute for Astronomy, School of Physics, A28, University of Sydney,
NSW, 2006, Australia}

\begin{abstract}
Gravitational lensing allows us to probe the structure of matter on a broad range
of astronomical scales, and as light from a distant source traverses 
an intervening galaxy, compact matter such as planets, stars,  and black holes
act as individual lenses. The magnification from such microlensing results in 
rapid brightness fluctuations which reveal not only the properties of the 
lensing masses, but also the surface brightness distribution in the source.
However, while the combination of deflections due to individual stars  is linear, the
resulting magnifications are highly non-linear, leading to significant computational
challenges which currently limit the range of problems which can be tackled.
This paper presents a new and novel implementation of a numerical approach 
to gravitational microlensing, increasing the scale of the problems that can be tackled 
by more than two orders of magnitude, opening up a new regime of 
astrophysically interesting problems.
\end{abstract}


\end{frontmatter}

\section{Introduction}\label{introduction}
\footnotetext[1]{Research undertaken as part of the Commonwealth Cosmology 
Initiative (CCI: www.thecci.org), an international collaboration 
supported by the Australian Research Council.}
Since the identification of the first multiply-imaged quasar \citep{walsh}, gravitational lensing
has been used to probe the distribution of matter on many astrophysical scales, such as planets
\citep{gaudi},
individual galaxies \citep[e.g.][]{2008MNRAS.390...39B,2008MNRAS.388..384D}, clusters
\citep[e.g.][]{2008ApJ...687...39D,2008ApJ...674..711S}, and large-scale structure 
\citep[e.g.][]{2008MNRAS.390..149K,2009ApJ...691..547W}.
Soon after the discovery of the first gravitational lens, it was realised that individual stars within a lensing
galaxy would themselves act as lenses
\citep{1979Natur.282..561C}, and while the separation of the resulting images would be too 
small to be resolved, this microlensing can significantly magnify the background source. 
Furthermore, given the density of compact objects within galaxies (such as
stars and possible dark matter), it was clear that many objects will influence the path of a beam of light
as it traverses a galaxy, and beyond considering only a couple of lensing masses, the study
of microlensing becomes analytically intractable and must be tackled numerically
\citep{1981ApJ...244..756Y,1986ApJ...301..503P}.

The first observations of microlensing were made in 1989 \citep{corrigan, irwin}, in  2237+0305,
a distant quasar (z = 1.695) lensed into four images by a nearby spiral galaxy \citep{huchra}; it 
has been photometrically monitored for two decades \citep{2006AcA....56..293U}, 
revealing exquisite light curves for each of the images, clearly revealing the 
presence of gravitational microlensing. Furthermore, the number of potentially 
microlensed quasars has steadily increased \citep{Eigenbrod, Gunn, Myers, Reimers, Turner, Vanderriest} and hence an efficient means of 
theoretically understanding gravitational microlensing is required.

\section{Microlensing Ray-Tracing}

\subsection{Mathematical Framework}
This paper does not intend to derive the basic gravitational lensing equations, 
the reader is directed towards the excellent textbook by \citet{1992grle.book.....S}  and thesis
by  \citet{wambsganss thesis}.
For microlensing calculations a normalized lens equation in the following form is used to
map the ray from the observer, through the lens, and into the source:
\begin{equation}
\vec{y} =
\left( \begin{array}{ccc}
1 - \gamma & 0 \\
0 & 1 + \gamma  \end{array} \right)\vec{x} - {\sigma}_c\vec{x} - \sum_{i=1}^{N_*}
\frac{m_i(\vec{x}-\vec{x_i})}{(\vec{x}-\vec{x_i})\textsuperscript{2}}.
\label{microlenseqn}
\end{equation}
Here, {$\vec{x}$} and {$\vec{y}$} are the location of a light ray in the lens and source plane respectively, {$\vec{x_i}$} and {$m_i$} are the mass and co-ordinates of the \emph{i}th microlensing star. For the remaining terms in this expression, $\sigma_c$ 
represents the local density of smooth matter in the galaxy, whereas $\gamma$, known as the shear, encapsulates the 
large scale asymmetry in the galaxy mass distribution.  The surface density of  compact objects (assumed to be uniform), 
is also used in modelling and simulations and is given by $\sigma_s$. The distance scale used
 in lensing 
is the Einstein Radius (ER), the radius of a ring produced by lensing of a source that is directly 
behind a point lens in a line from lens to observer. Einstein Radii are used to refer to the extent of observed images on the sky, and are derived from the mass of the lens and
the distances to the lens and source.

\begin{figure}[htp]
\centering
\includegraphics[height=58mm,width=58mm]{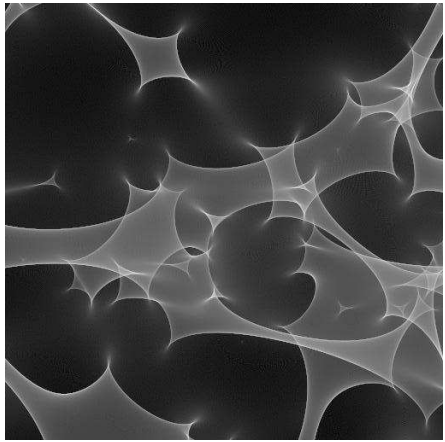}
\caption{An example of a gravitational microlensing magnification map generated via inverse
ray tracing. The lens consists of 435 stars each of 1 solar mass (M$_{\odot}$), randomly distributed, with mass parameters $\sigma_s$ = 0.39, $\gamma$ = 0.10. The viewing window
is 8 ER in width.
Light shading  corresponds to regions of high magnification, whereas 
darker regions represent demagnification.}\label{star_sims}
\end{figure}

\subsection{Numerical Approach}
Clearly, Eqn~\ref{microlenseqn} represents a mapping from a source location, $\vec{y}$, to a number of image (or observed)
positions $\vec{x}$. However, it is apparent that this mapping is one-to-many, with a single source resulting in a number
of images (dependent upon the number of microlensing masses $N_*$). However, reversing the mapping from 
the image plane back to the source is one-to-one and hence an ``inverse ray-tracing'' mechanism \citep{glassner} is typically employed in the
study of microlensing. With this, the observer sends out a large number of rays into the image plane. For each ray, the deflection
angle is calculated through Eqn~\ref{microlenseqn} and the ray is mapped into the source plane and collected on
a grid. After all the rays are followed, the density of rays in the  binned-up map of their source  positions is directly
proportional to the magnification of a source at that location; Fig~\ref{star_sims} presents an example of 
such a map, where light regions correspond to a high density of collected rays (i.e. strong magnification)
whereas dark regions correspond to the opposite situation.

\begin{figure}[htp]
\centering
\subfigure[100\% = 1\ M{$_\odot$}, number of stars = 30496]{
\includegraphics[height=58mm]{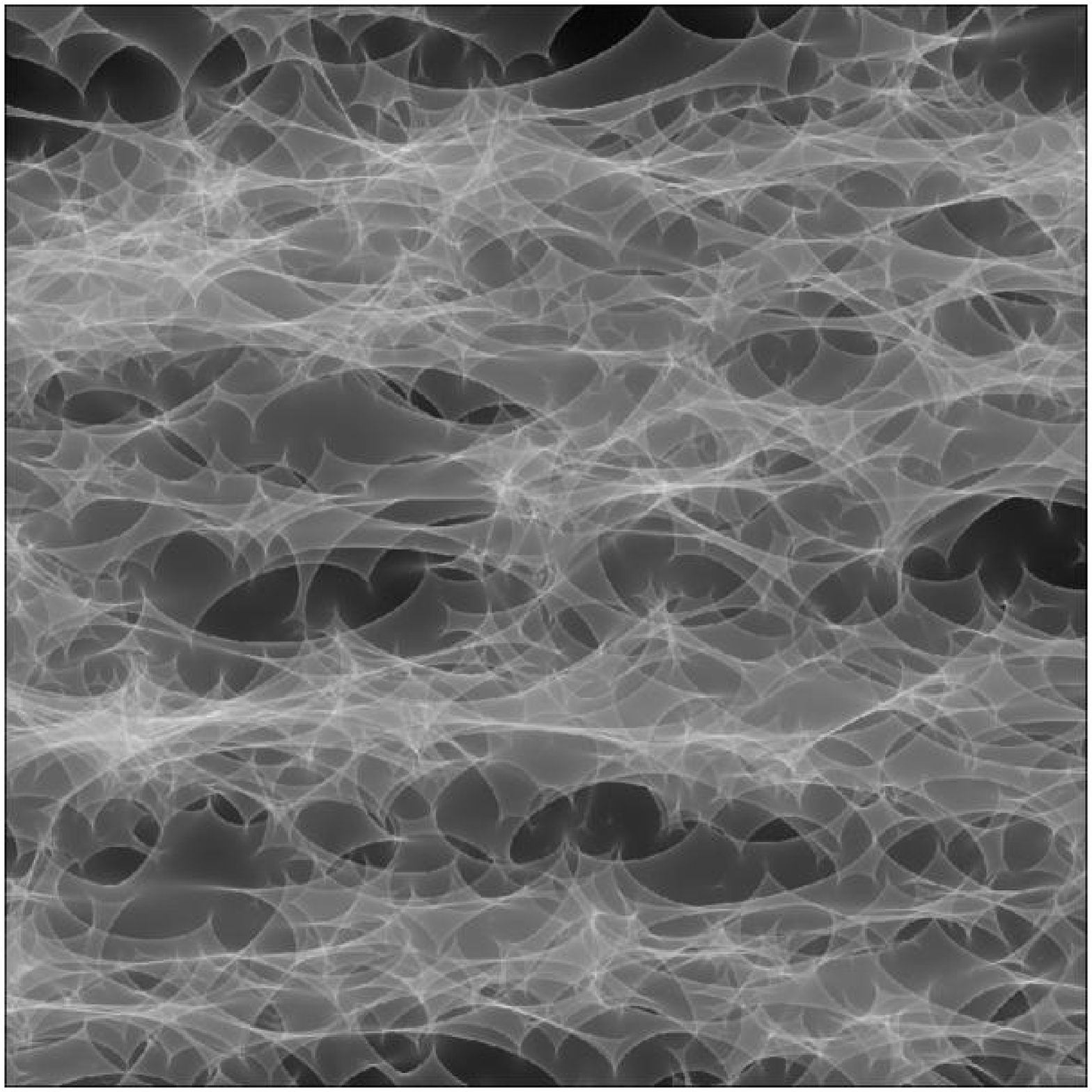}}
\subfigure[98\% = 0.01 M{$_\odot$}, number of stars = 2989306]{
\includegraphics[height=58mm]{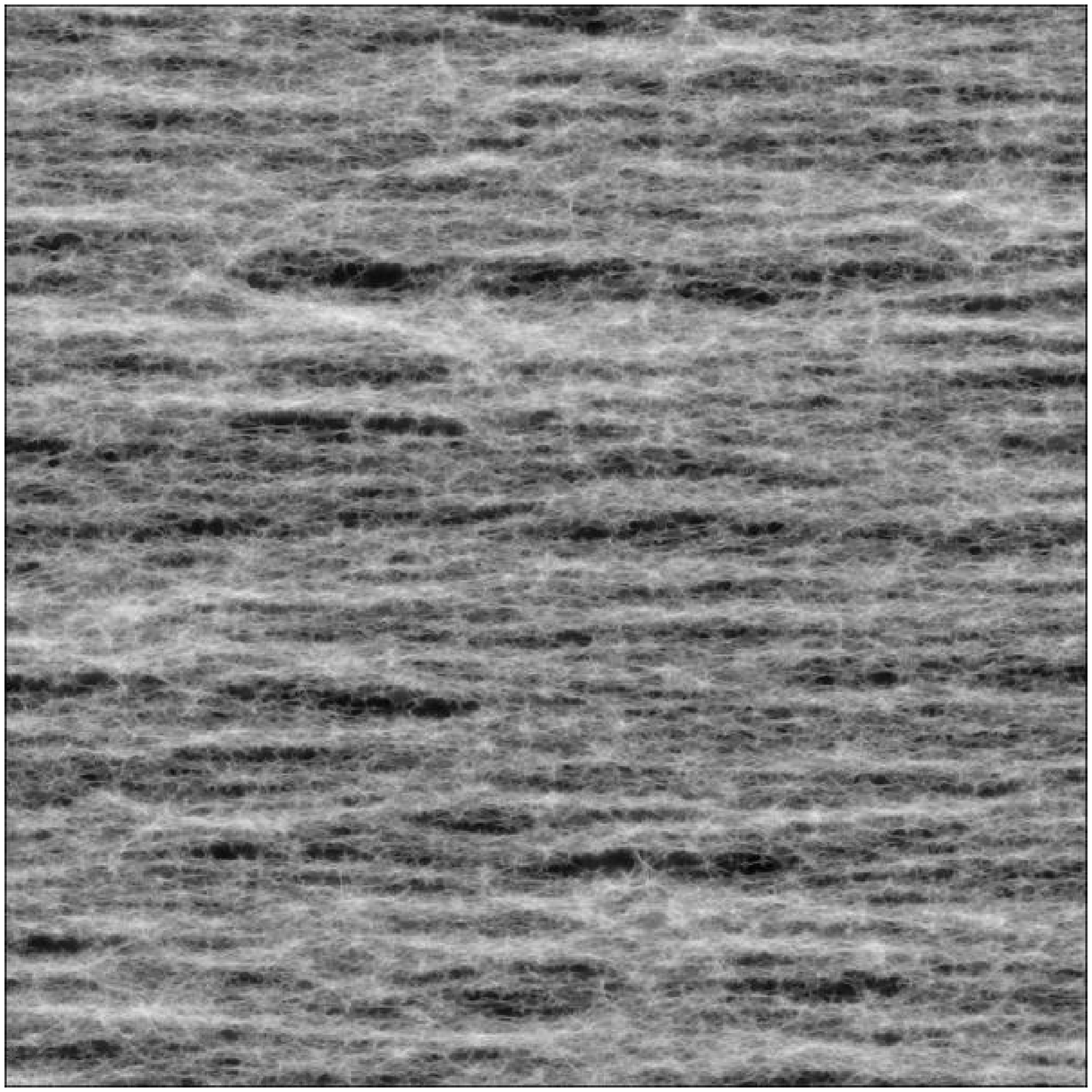}}
\subfigure[98\% = 0.001 M{$_\odot$}, number of stars = 29887581]{
\includegraphics[height=58mm]{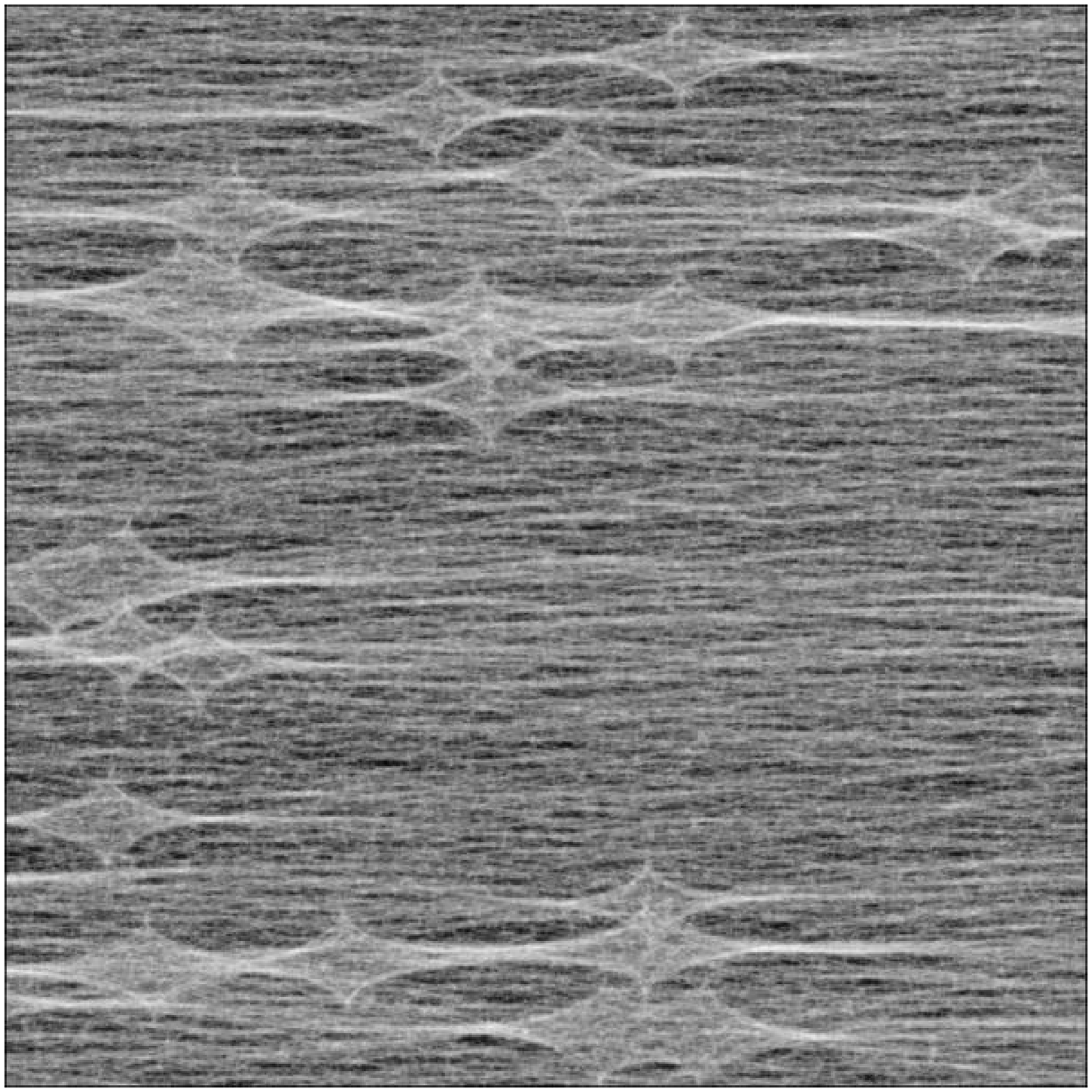}}
\subfigure[98\% = Smooth, number of stars = 609]{
\includegraphics[height=58mm]{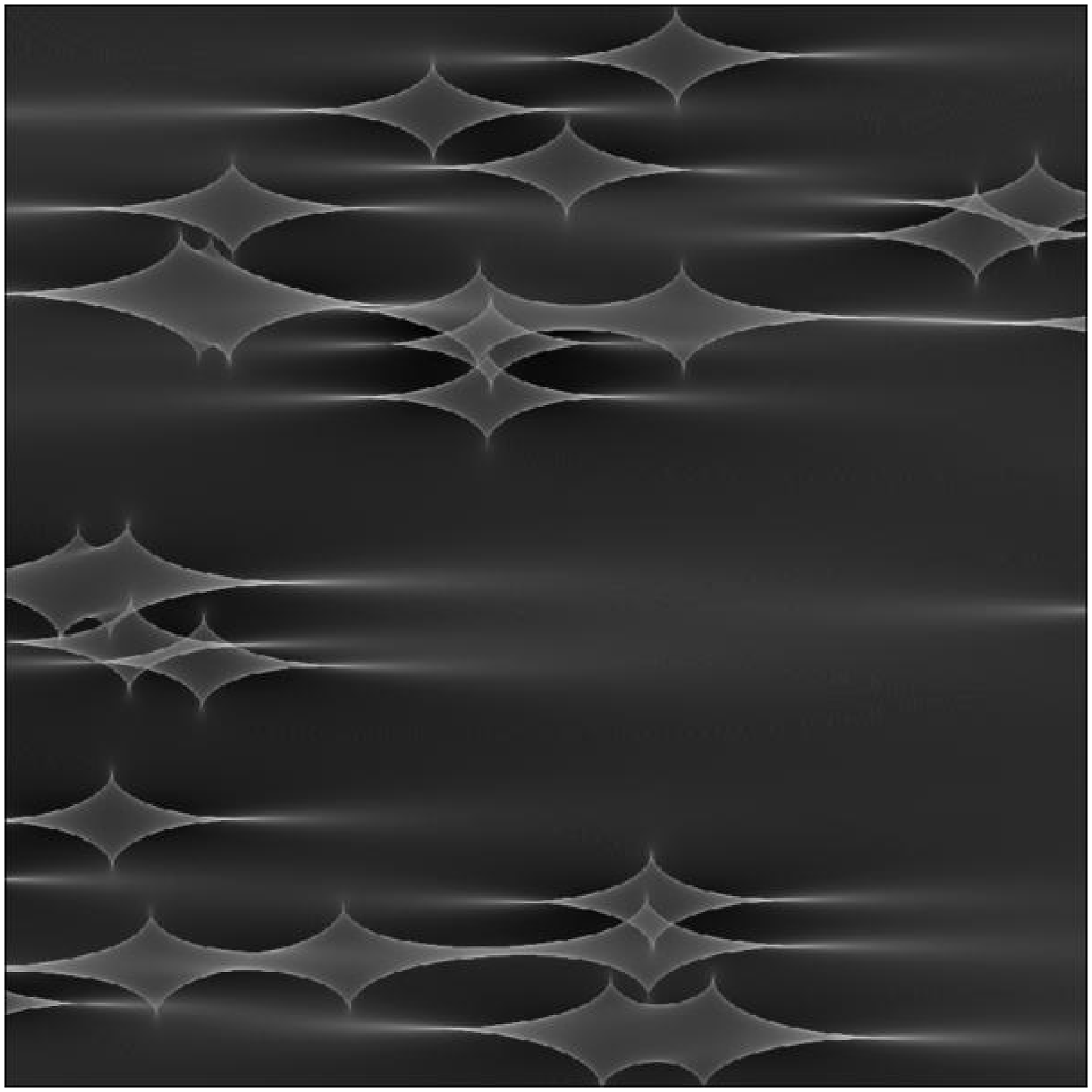}}
\caption{Magnification maps from simulations of bi-modal mass distributions.
A total mass of 30496 $M_{\odot}$ is distributed in a circular lens plane of diameter 500 ER, with a viewing window of 20 $\times$ 20 ER. All simulations use
$\gamma$ = 0.4250, with (a), (b) and (c) having $\sigma_s$ = 0.4750, $\sigma_c$ = 0 and (d) having $\sigma_s$ = 0.0095, $\sigma_c$ = 0.4655, so the total mass is the same in all.  The first simulation (a) is a single-mass distribution of stellar masses, then in (b) and (c) some of the stellar mass is replaced with smaller compact objects, then in (d) that mass is replaced with smooth matter. 
}\label{mag_maps}
\end{figure}

A deeper examination of Eqn~\ref{microlenseqn} reveals that the computationally intensive aspect of 
undertaking such ray tracing is the sum over the microlensing masses when calculating the deflection angles; in 
typical simulations, there could be $10^5\rightarrow10^6$ masses, with the tracing of $\sim10^{11}$rays 
required to achieve sufficient density in the source plane. Following the seminal work by 
\citep{1986A&A...166...36K},  \citet{wambsganss_comp} implemented an inverse ray-tracing approach which 
has become the ``industry standard'', utilizing a tree-algorithm to ease the sum over the microlensing 
deflection angles. The advent of this approach has allowed the analysis of the magnification patterns of
microlensed quasars \citep{1992ApJ...386...19W}, the structure of quasar broad line emission regions
\citep{keeton, lewis4}, chromatic effects in microlensing \citep{1991AJ....102..864W}, the nature of dark matter in lensing galaxies \citep{lewis1, pooley, schechter1}, and the effect of source size on microlensing \citep{bate, mortonson}, among others.

\section{New Physical Challenges}\label{challenge}
Several quasar systems appear to possess anomalous flux ratios \citep{Blackburne, Eigenbrod, Ota, Pooley2006}, meaning that the observed image
brightnesses differ significantly from predictions drawn from gravitational lens models possessing 
mass distributions that are smooth on galactic scales. Two key hypotheses have been put forward to 
explain these observations; either these anomalous ratios are due to millilensing by $\sim 10^6 {\rm
M_\odot}$ clumps of dark matter in the halo of the lensing galaxy \citep{Chiba, Dalal, Metcalf, Mao}, or the quasars are microlensed
by stars embedded in an overall smooth dark matter distribution \citep{witt}. The former proposal is 
attractive as it may provide the first direct measurement of the missing substructure expected 
from galactic build-up in cold dark matter formation scenarios. The latter proposal is also important, 
potentially probing the fundamental nature of dark matter, as will be explained below.

The original proposal by \citet{witt} considered a smooth dark matter component which actually suppresses
the image flux for long periods, leading to the apparently anomalous flux ratios. However, 
\citet{schechter1} questioned this  hypothesis, suggesting that the dark matter could in fact
be composed of substellar compact objects, and that the existence of such a compact 
component could imprint itself on the resulting gravitational lens statistics.
In studying the microlensing hypothesis, a number of of numerical simulations were undertaken
by \citet{schechter1} using the backwards ray shooting approach, and Figure \ref{mag_maps} illustrates 
several examples of the parameters employed. Figure \ref{mag_maps}(a) simulates a galaxy of a few
 thousand stars all of one solar mass (M{$_\odot$}), with $\sigma_s$ = 0.475 and $\gamma$ = 0.425. 
Figures \ref{mag_maps}(b) and (c) use the same parameters but two different masses
for $\sigma_s$: 2\% of the mass is in stars of mass M{$_\odot$}, and the rest is contained in 
small compact objects, 0.01M{$_\odot$} in  (b)  and 0.001M{$_\odot$} objects in  (c). 
The final panel, Figure \ref{mag_maps}(d), presents the case of a smooth dark matter component 
($\sigma_s$ = 0.0095, $\sigma_c$ = 0.4655, $\gamma$ = 0.425); as shown in \citep{lewis1}, 
when the mass of the compact dark matter component is made smaller, the general scale of caustic 
structure is reduced, although large scale caustic features, due to the presence of the solar mass
stars in the simulations, remain. In comparing the lower two panels in Figure~\ref{mag_maps}, the 
large scale caustic structure is the same in the small compact and smooth dark matter cases, but the
smaller scale structure is quite different (in fact, it is non-existent in the smooth matter case). 
\citet{lewis1} went on to show that, even though the large scale caustic structure is similar, 
the presence of the small scale caustics imprints itself on the statistical properties of the microlensing, 
with the lower panels possessing quite different magnification probability distributions. However, 
convolving these maps with a finite source radius washes out the small scale caustic structure, 
and for large enough sources the convolved compact mass and smooth mass magnification 
maps become statistically indistinguishable. This implies that there is a source scale size below
which any compact matter component would appear as compact, and above which it would appear
smooth, and hence observations of such systems at differing wavelengths (corresponding to differing 
source sizes) would probe the fundamental properties of dark matter.

However, the simulations undertaken to date have been limited by the computational power and 
memory of available computers; anomalous flux quasars are those that typically have been strongly
magnified, requiring ray shooting through a large area to be statistically meaningful. Furthermore, 
these limitations place constraints on the number of dark matter lenses that can be considered, 
effectively placing a limit on the mass scale that can be simulated  - around 10\textsuperscript{7}
objects of mass 0.001 M${\odot}$. Hence, it was decided that 
a supercomputer implementation of the ray-shooting approach was required to tackle these
physically interesting regimes.

\section{A Supercomputing Approach}
The  numerical ray-shooting approach employs several approximations that have little impact on the accuracy but allow a significant improvement in performance. 
An examination of Equation~\ref{microlenseqn}
reveals that a $r^{-1}$ calculation is required for each lensing mass,  the total deflection 
requiring a sum over all of these individual masses; of course, the calculation of the distance, $r$, involves the square roots, significantly adding to the computational load. The ray-tracing approach
as adopted by \citet{wambsganss_comp} reduces this load by use of a multipole expansion of the
gravitational interaction for lensing masses far from the passage of the ray through the field of 
masses. For this, a cell tree is constructed by subdividing the distribution of lensing masses, 
with a constant regular subdivision until each cell holds a maximum of a single lensing mass.
An opening angle criteria is then applied to distinguish between employing nearby masses 
as individual lenses, whereas more distant lens masses are grouped and a multipole expansion of their 
distribution is used to approximate their gravitational influences.

The use of the cell-tree substantially improves simulation run-time but substantially increases the amount of data required to run a simulation. All data is held in computer memory, and consists of:
\begin{itemize}
\item A 2-D array for the pixel map.
\item An array of ``stars'',  each element containing the mass and location of a star in the galaxy.
We will use the term ``stars'' to refer to all objects constituting  the lens.
\item An array for the cell tree, each element containing  information about a cell and its location within the tree.
\end{itemize}

A typical simulation on a desktop computer may contain a pixel array of 2000{$\times$}2000 pixels and 
of order 
a million stars. This corresponds to 15 Mb for the pixel array, 38 Mb for the stars array, and 89Mb for the cell tree. The rays have no data existence; the result of shooting a ray is a pixel hit which is added to that pixel's hit count. In the following, we outline the basic operation of the standard microlensing approach
and the changes required to employ it in a supercomputer environment, successfully handling the
significant computational and storage requirements to push the application of microlensing into 
a new scientific regime (see Section~\ref{results}).

The starting point is to understand the operation of the standard ray shooting approach of
\citet{wambsganss_comp}; this is expressed in the following 
 pseudo-code\begin{quote}
// {\bf Setup phase}

Generate the stars, using a random distribution of locations and masses, within certain parameters 

Sort the stars by encoding indexes

Generate the cell tree from the stars

Sort the cell tree by encoding indexes

// {\bf Shooting phase}

Loop
	\begin{quote}
	Shoot a ray\\
	Increment the pixel hit array where necessary\\
	Move to a new location in the lens plane
	\end{quote}
Until the whole lens plane has been covered
\end{quote}
The sorting operations of the stars and cell tree are required because during the shooting of the rays, searches are performed to determine the break between the employment of individual masses and  
expanded cells that are to be used for a particular ray, with the configuration changing as we 
consider rays traversing differing locations in the field of stars.

As noted in Section~\ref{challenge}, we wish to increase the number of stars under consideration. 
Immediately, this significantly impacts the undertaking of microlensing simulations by desktop 
computers (which are limited by virtual memory to  $\sim 20$ million stars), 
increasing the storage required for the masses and locations, as well as the tree structure.
Furthermore, we additionally require an increase of the number of rays fired; as can be seen in 
Figure~\ref{mag_maps}, the inclusion of small masses induces small scale structure in the magnification map, and 
to fully resolve this it is essential to increase the number of pixels covering the source, and the 
number of rays traced, to ensure we are not limited by counting statistics in the final map.
In moving to a supercomputer environment, the following changes are required:
\begin{itemize}
\item Adapt the  algorithmic approach of \citet{wambsganss_comp}.
\item Parallelize the approach so it can exploit current multiprocessor computers; ray tracing is inherently parallelizable, as the rays are usually independent of each other. There are typically billions of rays in a microlensing simulation, which can be distributed in some appropriate way across many processors and shot concurrently.
\item Increase the number of stars that can be considered, into the billions. This is an increase of a hundred-fold, and it is unlikely the data can be held in RAM with any reasonable architecture. Other strategies will be required to store the massive amounts of data that will be used.
\item Change to an object-oriented paradigm; the approach of \citet{wambsganss_comp} was originally implemented in Fortran 77 with procedural code and global data. It was decided that the supercomputer implementation would employ  C++ as some objects such as the stars and cells naturally lend themselves to implementation as C++ class objects. Re-implementing in C++ will provide a sound basis for future work.
\end{itemize}
As the reimplementation in C++ is a design and development process we will  describe it only briefly here.
Data elements, such as stars and cells, were identified and reimplemented as C++ class types.  The Fortran lensing algorithms were rewritten in C++ and combined with the C++ classes,
producing a working program. The arrays that store the stars and cells were replaced with
a C++ module that implements arrays as disk files (described later),  but with a similar interface. Then the implementation
was  parallelized
 by sharing out the generation of stars, and the ray-shooting, to different processors (this is described later). 

The rest of the paper focuses on the strategies used to parallelize our approach and increase the size of lenses 
available for
 microlensing research.

\subsubsection{Parallelizing the ray shooting}
For the supercomputer implementation, 
the rays that are shot need to be distributed across the available processors. There are many ways of parallelizing ray-tracing \citep{amri,lee,verdu} but little experience with this in gravitational lensing simulations. It does not matter which ray is executed on which processor, or in what order, as they are completely independent. 
We consider only static assignment of rays to processors, as dynamic assignment for load-balancing purposes is much more complicated and static assignment should be tried first. Dynamic load balancing is a model where job tasks (in this case rays)
are assigned to processors at run-time based on the load on each processor as the simulation proceeds - idle processors get the
next jobs, much like a multiprocessor operating system running user programs. Much more time and effort
is required to implement dynamic load balancing
compared to static load balancing. Static load balancing means that when the program runs, some algorithm
at startup determines which ray is assigned to which process, and that is fixed for the duration of the simulation.
 
 Therefore we are looking for a mapping 
\begin{equation}
Assign(Ray) \mapsto Processor
\end{equation}
 which maps rays to processors for that run. There is a value for \emph{Assign} which is optimal for a particular simulation, but it is unknown, so some heuristic is required. Mapping an equal number of rays to each processor is sufficient if there are many more rays than processors and each ray takes the same time. The latter is probably not going to be the case since the stars in the lens may not be not distributed evenly, and rays close to a lot of mass will take longer to process. We have adopted an interlaced ray distribution model: take a 2-D element \emph{dA} of the lens plane and distribute the rays in that element across all processors, and do the same for all other elements. This means that every processor simulates lensing over the entire lens plane, but only some rays at each point of the lens plane. We hope that this will even out the load without the need for
  load balancing. 

To implement parallelization  we could employ either a multithreading or multiprocess model. Multiprocessing means that several parallel processes are created by the computer operating system and they must access shared data  outside the program,
for example on disk. In multithreading there is only a single process with parallel threads of execution running inside it, and all global data in
the program is automatically shared between threads. Multithreading is 
usually more efficient but also more error-prone in development, so for a first
implementation we opted for multiprocessing.

 In the setup phase the generation of the stars can be parallelized by distributing the load evenly across all processes.  The generation of the cells from the stars is a sequential process based on sophisticated location and tree indexing which cannot be easily parallelized. The sorting steps could be parallelized but have not been; this will be discussed in the next section.
The shooting phase can be distributed across all processors.

\subsection{Increasing the data storage}
Pushing into the regime of $\sim$billion stars in a lens requires a stars array and cell
tree of order 200GB; the following are possible ways of storing this data:
\begin{enumerate}
\item On a 64-bit computer with 200Gb RAM it could be stored in memory as it previously has been.
\item If simulations are to be run on many computers at the same time, and the computers collectively have enough memory, distribute the data in the memory of those machines.
\item Put the data in files on a shared disk, not in RAM.
\end{enumerate}
Option 1 is possible and requires no modification of the data storage, although
clearly requires a particular computer architecture not available in typical (i.e. beowulf) clusters.
Option 2 is  possible if enough machines are used with enough RAM; then data could be spread amongst them and fetched across the network. Option 3 can be used anywhere there is sufficient disk space, even on desktop computers. In our new implementation all three options were combined, as huge amounts of disk space are now common (more than large RAM machines), the simulation data is placed on disk, but partially cached in memory so that all the available RAM is used. 

The stars array  and cell tree were each replaced with a binary  file, but still accessed like arrays, so that any object at any part of the file could be requested with an index. Naturally this means that file I/O will dramatically slow down the simulation, although it is envisaged that caching strategies and parallelizing the ray-shooting may compensate for this. 
Since shared files are trivial on modern networks,  multiple processes can read data from a single set of shared files. When starting a simulation,  several processes can create the files and the others wait until they  are ready, or,  each process can generate its own version of the files, which is very wasteful of disk space but means some processes may be able to start earlier.

\subsection{Sorting with files}
The sorting of stars and cells is inherent in the algorithms in the set-up phase that code the stars and cells for indexing and searching; it cannot be avoided without significant redesign of the cell tree usage. In the original implementation where data is in RAM, a heap-sort algorithm is used \citep{knuth}. A heap sort swaps elements from the end of an array to the beginning, gradually 
building a sorted array in place of the original values. When the stars and cell tree were placed in files the sorting steps became extremely slow - as expected. A run of one billion stars would take an intolerably long time and be unworkable. In fact, when the data was initially placed in files and a simulation of one billion attempted, the sort was still running after two days and we gave up on it.

The best way to process files is sequentially, and best way to sort files is with a merge sort \citep{knuth}. We implemented a folding merge-sort algorithm for the sorting steps. In the original implementation the stars and cell tree were sorted after they were generated; a folding merge-sort does the sorting \emph{while} the data is generated.  The algorithm is as follows; as stars are generated they are placed in a memory cache. When the cache is full, it is sorted using a heap-sort and written to the stars file. This is the first instance of the file. Further stars are generated, and  placed in the cache. When the cache is full it is heap-sorted and then merge-sorted into the file, producing a new instance of the file. The process continues until all stars are in the file, and the file is always sorted. The same process is done for the cell tree.

This mechanism substantially improved the run-time of the sorting phases. There are however disadvantages:
\begin{itemize}
\item The space required for a merge-sort is double the data size, due  to the fact that a new version of the file is being created as the existing one is being read
\item As the file gets bigger the merge gets slower 
\end{itemize}
Nevertheless this method has provided us with a usable implementation.

\subsection{Shooting cache}
The memory cache used for merge-sorting is only required during the setup phase; after that the memory is available for shooting. As rays are shot, a per-ray stars and cell configuration is used to calculate the deflection, and analysis of the usage patterns in the sequential version showed that around 66\% of stars and cells used for one ray are used again for the next, as the rays are close together in space (assuming a large number of them).

Based on this information we implemented a shooting memory cache that allows re-use of stars and cells, from one ray to the next. Unfortunately this scheme interacts with the interlaced ray distribution scheme, in that the scheme means  each process is executing rays that are slightly further apart on the lens plane than they normally would be.  This degrades the performance of the shooting cache slightly. 

\subsection{File cache}
The amount of memory needed for the stars/cell configuration for a ray is not much (a few Mb), leaving plenty of RAM free. This RAM could be used as a more general memory cache for storing some stars and cells that are otherwise on disk, making
them faster to access. Implementing a general  cache of file data requires an analysis of the usage patterns of the stars and cell data files, and at a first-order approximation there are two patterns operating: depth first traversals  of the cell tree and binary searches in both the stars and cell arrays. A depth
first traversal will follow a single branch of the cell tree from the root down
to  a "leaf", traversing the file in a sequential fashion; a binary search
will jump around from one end of the file to the other. These conflict, and make it difficult to generate heuristics for anticipating what portions of the files to load into memory, so we have chosen a simple scheme to begin with: cache the first  levels of the cell tree and the first records of the stars file, as many as possible. The root of the cell tree is heavily accessed and caching it proved useful. 

\section{Results}\label{results}
We discuss the performance of our new method and then give some preliminary results of its application to microlensing.

\subsection{Performance}
The following performance statistics were obtained from simulations using 10\textsuperscript{7} stars with a pixel map of 2000x2000. The number of rays to shoot are based on these and other parameters, in this case it is 1.3x10\textsuperscript{9}. The performance  depends greatly on the amount of RAM available for caching. For the test runs reported here, we fixed the size of the cache per parallel process to be about 1/8th the size of the stars data, because this is the ratio for a simulation of 10\textsuperscript{9} stars run on a common 32-bit desktop machine, something we may consider as a ``standard'' for comparisons. 


\begin{table}
\centering
\begin{tabular}{ | l | c | c | c | c |}
\hline
& {\bf Version} & {\bf Total} & {\bf Shooting} & {\bf Setup} \\
& & {\bf time}  & {\bf time} & {\bf time}  \\
\hline
{\bf Fortran} &   v1 & 5.64  &      5.57  &    0.06    \\
\hline
{\bf Fortran, optimized} & v2 &   1.98    &    1.93   &   0.05    \\
\hline
{\bf C++, Fortran copy} & v3 & 5.38   &     5.34   &   0.04 \\
\hline
{\bf C++, copy, optimized} & v4 & 2.15    &    2.12   &   0.03  \\
\hline
{\bf C++, data in files,} & v5 & 20.93   &    19.72   &   1.21      \\
{\bf no caches} & & & & \\
\hline
{\bf C++, add merge-sorts} & v6 &  4.20   &     4.01   &   0.18        \\
{\bf and file caches} & & & & \\
\hline
\end{tabular}
\caption{Run times (hours) for different versions of the new approach, single process}
\label{program times}
\end{table}

Table \ref{program times} shows run-times for different versions of the new tool, beginning with the original Fortran (v1) and progressing through some C++ versions to the final version (v6). All times are in hours.
The first C++ version that was generated (v3, v4)  was a straight copy of the Fortran (v1, v2). The run-times for both languages are similar. Of note is the fact that both  run  2.5-2.8 times faster when full compiler optimizations are used, due to the algorithms being computationally intensive and these versions having no I/O operations.
The C++ is slightly computationally faster when unoptimized, the Fortran slightly faster when optimized, but they are close, so we have not lost much in the move to C++. The same compilers (GNU C++ and GNU Fortran 95 from FSF\footnote{Free Software Foundation, Inc.
    51 Franklin St, Fifth Floor, Boston, MA 02110-1301 USA}) were used
for both languages, but since there are overheads in C++ due to the use of object-oriented classes, it
is not surprising that C++ is slightly slower in the best-case scenario. We have not investigated
the relative performance when both are unoptimized. Different compilers with
different optimization strategies can always produce a faster (or slower) program, so our comparisons do not
constitute a formal benchmarch, merely the performance for commonly-used compilers (i.e GNU). Further investigation can be made in the future to determine the compiler that is best suited for this program.

After taking the arrays out of memory and placing them on disk the performance, as expected,  degrades considerably (v5).  The run is 10 times slower. Although the stars and cells are generated in a sequential fashion, they are then heap-sorted, which is very inefficient inside files. Using the folding merge-sort (v6) avoids inefficiencies because it scans and merges file sections sequentially. In v5 when shooting begins there is no file caching; star data and cell data is loaded from all over the files, and the I/O and spread-out file access makes shooting very slow. In v6 the shooting cache and file cache are implemented and they improve performance time by reducing the access to the star and cell data files. The techniques used to improve performance in the files version (v6) bring the run-time down so that is is comparable to the unoptimized in-memory version  - a good result.

\subsubsection{Parallel speed-up}
Versions 4 and 6 are able to run in parallel mode, dividing work between processes. Version 4 maintains data in memory and is can be used when the number of stars is up to 10\textsuperscript{7}; version 6 uses files and must be used for simulations of more than 10\textsuperscript{7} stars.

\begin{figure}
\centering
\begingroup%
  \makeatletter%
  \newcommand{\GNUPLOTspecial}{%
    \@sanitize\catcode`\%=14\relax\special}%
  \setlength{\unitlength}{0.1bp}%
\begin{picture}(3600,2160)(0,0)%
{\GNUPLOTspecial{"
/gnudict 256 dict def
gnudict begin
/Color false def
/Solid false def
/gnulinewidth 5.000 def
/userlinewidth gnulinewidth def
/vshift -33 def
/dl {10.0 mul} def
/hpt_ 31.5 def
/vpt_ 31.5 def
/hpt hpt_ def
/vpt vpt_ def
/Rounded false def
/M {moveto} bind def
/L {lineto} bind def
/R {rmoveto} bind def
/V {rlineto} bind def
/N {newpath moveto} bind def
/C {setrgbcolor} bind def
/f {rlineto fill} bind def
/vpt2 vpt 2 mul def
/hpt2 hpt 2 mul def
/Lshow { currentpoint stroke M
  0 vshift R show } def
/Rshow { currentpoint stroke M
  dup stringwidth pop neg vshift R show } def
/Cshow { currentpoint stroke M
  dup stringwidth pop -2 div vshift R show } def
/UP { dup vpt_ mul /vpt exch def hpt_ mul /hpt exch def
  /hpt2 hpt 2 mul def /vpt2 vpt 2 mul def } def
/DL { Color {setrgbcolor Solid {pop []} if 0 setdash }
 {pop pop pop 0 setgray Solid {pop []} if 0 setdash} ifelse } def
/BL { stroke userlinewidth 2 mul setlinewidth
      Rounded { 1 setlinejoin 1 setlinecap } if } def
/AL { stroke userlinewidth 2 div setlinewidth
      Rounded { 1 setlinejoin 1 setlinecap } if } def
/UL { dup gnulinewidth mul /userlinewidth exch def
      dup 1 lt {pop 1} if 10 mul /udl exch def } def
/PL { stroke userlinewidth setlinewidth
      Rounded { 1 setlinejoin 1 setlinecap } if } def
/LTw { PL [] 1 setgray } def
/LTb { BL [] 0 0 0 DL } def
/LTa { AL [1 udl mul 2 udl mul] 0 setdash 0 0 0 setrgbcolor } def
/LT0 { PL [] 1 0 0 DL } def
/LT1 { PL [4 dl 2 dl] 0 1 0 DL } def
/LT2 { PL [2 dl 3 dl] 0 0 1 DL } def
/LT3 { PL [1 dl 1.5 dl] 1 0 1 DL } def
/LT4 { PL [5 dl 2 dl 1 dl 2 dl] 0 1 1 DL } def
/LT5 { PL [4 dl 3 dl 1 dl 3 dl] 1 1 0 DL } def
/LT6 { PL [2 dl 2 dl 2 dl 4 dl] 0 0 0 DL } def
/LT7 { PL [2 dl 2 dl 2 dl 2 dl 2 dl 4 dl] 1 0.3 0 DL } def
/LT8 { PL [2 dl 2 dl 2 dl 2 dl 2 dl 2 dl 2 dl 4 dl] 0.5 0.5 0.5 DL } def
/Pnt { stroke [] 0 setdash
   gsave 1 setlinecap M 0 0 V stroke grestore } def
/Dia { stroke [] 0 setdash 2 copy vpt add M
  hpt neg vpt neg V hpt vpt neg V
  hpt vpt V hpt neg vpt V closepath stroke
  Pnt } def
/Pls { stroke [] 0 setdash vpt sub M 0 vpt2 V
  currentpoint stroke M
  hpt neg vpt neg R hpt2 0 V stroke
  } def
/Box { stroke [] 0 setdash 2 copy exch hpt sub exch vpt add M
  0 vpt2 neg V hpt2 0 V 0 vpt2 V
  hpt2 neg 0 V closepath stroke
  Pnt } def
/Crs { stroke [] 0 setdash exch hpt sub exch vpt add M
  hpt2 vpt2 neg V currentpoint stroke M
  hpt2 neg 0 R hpt2 vpt2 V stroke } def
/TriU { stroke [] 0 setdash 2 copy vpt 1.12 mul add M
  hpt neg vpt -1.62 mul V
  hpt 2 mul 0 V
  hpt neg vpt 1.62 mul V closepath stroke
  Pnt  } def
/Star { 2 copy Pls Crs } def
/BoxF { stroke [] 0 setdash exch hpt sub exch vpt add M
  0 vpt2 neg V  hpt2 0 V  0 vpt2 V
  hpt2 neg 0 V  closepath fill } def
/TriUF { stroke [] 0 setdash vpt 1.12 mul add M
  hpt neg vpt -1.62 mul V
  hpt 2 mul 0 V
  hpt neg vpt 1.62 mul V closepath fill } def
/TriD { stroke [] 0 setdash 2 copy vpt 1.12 mul sub M
  hpt neg vpt 1.62 mul V
  hpt 2 mul 0 V
  hpt neg vpt -1.62 mul V closepath stroke
  Pnt  } def
/TriDF { stroke [] 0 setdash vpt 1.12 mul sub M
  hpt neg vpt 1.62 mul V
  hpt 2 mul 0 V
  hpt neg vpt -1.62 mul V closepath fill} def
/DiaF { stroke [] 0 setdash vpt add M
  hpt neg vpt neg V hpt vpt neg V
  hpt vpt V hpt neg vpt V closepath fill } def
/Pent { stroke [] 0 setdash 2 copy gsave
  translate 0 hpt M 4 {72 rotate 0 hpt L} repeat
  closepath stroke grestore Pnt } def
/PentF { stroke [] 0 setdash gsave
  translate 0 hpt M 4 {72 rotate 0 hpt L} repeat
  closepath fill grestore } def
/Circle { stroke [] 0 setdash 2 copy
  hpt 0 360 arc stroke Pnt } def
/CircleF { stroke [] 0 setdash hpt 0 360 arc fill } def
/C0 { BL [] 0 setdash 2 copy moveto vpt 90 450  arc } bind def
/C1 { BL [] 0 setdash 2 copy        moveto
       2 copy  vpt 0 90 arc closepath fill
               vpt 0 360 arc closepath } bind def
/C2 { BL [] 0 setdash 2 copy moveto
       2 copy  vpt 90 180 arc closepath fill
               vpt 0 360 arc closepath } bind def
/C3 { BL [] 0 setdash 2 copy moveto
       2 copy  vpt 0 180 arc closepath fill
               vpt 0 360 arc closepath } bind def
/C4 { BL [] 0 setdash 2 copy moveto
       2 copy  vpt 180 270 arc closepath fill
               vpt 0 360 arc closepath } bind def
/C5 { BL [] 0 setdash 2 copy moveto
       2 copy  vpt 0 90 arc
       2 copy moveto
       2 copy  vpt 180 270 arc closepath fill
               vpt 0 360 arc } bind def
/C6 { BL [] 0 setdash 2 copy moveto
      2 copy  vpt 90 270 arc closepath fill
              vpt 0 360 arc closepath } bind def
/C7 { BL [] 0 setdash 2 copy moveto
      2 copy  vpt 0 270 arc closepath fill
              vpt 0 360 arc closepath } bind def
/C8 { BL [] 0 setdash 2 copy moveto
      2 copy vpt 270 360 arc closepath fill
              vpt 0 360 arc closepath } bind def
/C9 { BL [] 0 setdash 2 copy moveto
      2 copy  vpt 270 450 arc closepath fill
              vpt 0 360 arc closepath } bind def
/C10 { BL [] 0 setdash 2 copy 2 copy moveto vpt 270 360 arc closepath fill
       2 copy moveto
       2 copy vpt 90 180 arc closepath fill
               vpt 0 360 arc closepath } bind def
/C11 { BL [] 0 setdash 2 copy moveto
       2 copy  vpt 0 180 arc closepath fill
       2 copy moveto
       2 copy  vpt 270 360 arc closepath fill
               vpt 0 360 arc closepath } bind def
/C12 { BL [] 0 setdash 2 copy moveto
       2 copy  vpt 180 360 arc closepath fill
               vpt 0 360 arc closepath } bind def
/C13 { BL [] 0 setdash  2 copy moveto
       2 copy  vpt 0 90 arc closepath fill
       2 copy moveto
       2 copy  vpt 180 360 arc closepath fill
               vpt 0 360 arc closepath } bind def
/C14 { BL [] 0 setdash 2 copy moveto
       2 copy  vpt 90 360 arc closepath fill
               vpt 0 360 arc } bind def
/C15 { BL [] 0 setdash 2 copy vpt 0 360 arc closepath fill
               vpt 0 360 arc closepath } bind def
/Rec   { newpath 4 2 roll moveto 1 index 0 rlineto 0 exch rlineto
       neg 0 rlineto closepath } bind def
/Square { dup Rec } bind def
/Bsquare { vpt sub exch vpt sub exch vpt2 Square } bind def
/S0 { BL [] 0 setdash 2 copy moveto 0 vpt rlineto BL Bsquare } bind def
/S1 { BL [] 0 setdash 2 copy vpt Square fill Bsquare } bind def
/S2 { BL [] 0 setdash 2 copy exch vpt sub exch vpt Square fill Bsquare } bind def
/S3 { BL [] 0 setdash 2 copy exch vpt sub exch vpt2 vpt Rec fill Bsquare } bind def
/S4 { BL [] 0 setdash 2 copy exch vpt sub exch vpt sub vpt Square fill Bsquare } bind def
/S5 { BL [] 0 setdash 2 copy 2 copy vpt Square fill
       exch vpt sub exch vpt sub vpt Square fill Bsquare } bind def
/S6 { BL [] 0 setdash 2 copy exch vpt sub exch vpt sub vpt vpt2 Rec fill Bsquare } bind def
/S7 { BL [] 0 setdash 2 copy exch vpt sub exch vpt sub vpt vpt2 Rec fill
       2 copy vpt Square fill
       Bsquare } bind def
/S8 { BL [] 0 setdash 2 copy vpt sub vpt Square fill Bsquare } bind def
/S9 { BL [] 0 setdash 2 copy vpt sub vpt vpt2 Rec fill Bsquare } bind def
/S10 { BL [] 0 setdash 2 copy vpt sub vpt Square fill 2 copy exch vpt sub exch vpt Square fill
       Bsquare } bind def
/S11 { BL [] 0 setdash 2 copy vpt sub vpt Square fill 2 copy exch vpt sub exch vpt2 vpt Rec fill
       Bsquare } bind def
/S12 { BL [] 0 setdash 2 copy exch vpt sub exch vpt sub vpt2 vpt Rec fill Bsquare } bind def
/S13 { BL [] 0 setdash 2 copy exch vpt sub exch vpt sub vpt2 vpt Rec fill
       2 copy vpt Square fill Bsquare } bind def
/S14 { BL [] 0 setdash 2 copy exch vpt sub exch vpt sub vpt2 vpt Rec fill
       2 copy exch vpt sub exch vpt Square fill Bsquare } bind def
/S15 { BL [] 0 setdash 2 copy Bsquare fill Bsquare } bind def
/D0 { gsave translate 45 rotate 0 0 S0 stroke grestore } bind def
/D1 { gsave translate 45 rotate 0 0 S1 stroke grestore } bind def
/D2 { gsave translate 45 rotate 0 0 S2 stroke grestore } bind def
/D3 { gsave translate 45 rotate 0 0 S3 stroke grestore } bind def
/D4 { gsave translate 45 rotate 0 0 S4 stroke grestore } bind def
/D5 { gsave translate 45 rotate 0 0 S5 stroke grestore } bind def
/D6 { gsave translate 45 rotate 0 0 S6 stroke grestore } bind def
/D7 { gsave translate 45 rotate 0 0 S7 stroke grestore } bind def
/D8 { gsave translate 45 rotate 0 0 S8 stroke grestore } bind def
/D9 { gsave translate 45 rotate 0 0 S9 stroke grestore } bind def
/D10 { gsave translate 45 rotate 0 0 S10 stroke grestore } bind def
/D11 { gsave translate 45 rotate 0 0 S11 stroke grestore } bind def
/D12 { gsave translate 45 rotate 0 0 S12 stroke grestore } bind def
/D13 { gsave translate 45 rotate 0 0 S13 stroke grestore } bind def
/D14 { gsave translate 45 rotate 0 0 S14 stroke grestore } bind def
/D15 { gsave translate 45 rotate 0 0 S15 stroke grestore } bind def
/DiaE { stroke [] 0 setdash vpt add M
  hpt neg vpt neg V hpt vpt neg V
  hpt vpt V hpt neg vpt V closepath stroke } def
/BoxE { stroke [] 0 setdash exch hpt sub exch vpt add M
  0 vpt2 neg V hpt2 0 V 0 vpt2 V
  hpt2 neg 0 V closepath stroke } def
/TriUE { stroke [] 0 setdash vpt 1.12 mul add M
  hpt neg vpt -1.62 mul V
  hpt 2 mul 0 V
  hpt neg vpt 1.62 mul V closepath stroke } def
/TriDE { stroke [] 0 setdash vpt 1.12 mul sub M
  hpt neg vpt 1.62 mul V
  hpt 2 mul 0 V
  hpt neg vpt -1.62 mul V closepath stroke } def
/PentE { stroke [] 0 setdash gsave
  translate 0 hpt M 4 {72 rotate 0 hpt L} repeat
  closepath stroke grestore } def
/CircE { stroke [] 0 setdash 
  hpt 0 360 arc stroke } def
/Opaque { gsave closepath 1 setgray fill grestore 0 setgray closepath } def
/DiaW { stroke [] 0 setdash vpt add M
  hpt neg vpt neg V hpt vpt neg V
  hpt vpt V hpt neg vpt V Opaque stroke } def
/BoxW { stroke [] 0 setdash exch hpt sub exch vpt add M
  0 vpt2 neg V hpt2 0 V 0 vpt2 V
  hpt2 neg 0 V Opaque stroke } def
/TriUW { stroke [] 0 setdash vpt 1.12 mul add M
  hpt neg vpt -1.62 mul V
  hpt 2 mul 0 V
  hpt neg vpt 1.62 mul V Opaque stroke } def
/TriDW { stroke [] 0 setdash vpt 1.12 mul sub M
  hpt neg vpt 1.62 mul V
  hpt 2 mul 0 V
  hpt neg vpt -1.62 mul V Opaque stroke } def
/PentW { stroke [] 0 setdash gsave
  translate 0 hpt M 4 {72 rotate 0 hpt L} repeat
  Opaque stroke grestore } def
/CircW { stroke [] 0 setdash 
  hpt 0 360 arc Opaque stroke } def
/BoxFill { gsave Rec 1 setgray fill grestore } def
/BoxColFill {
  gsave Rec
  /Fillden exch def
  currentrgbcolor
  /ColB exch def /ColG exch def /ColR exch def
  /ColR ColR Fillden mul Fillden sub 1 add def
  /ColG ColG Fillden mul Fillden sub 1 add def
  /ColB ColB Fillden mul Fillden sub 1 add def
  ColR ColG ColB setrgbcolor
  fill grestore } def
%
%
/PatternFill { gsave /PFa [ 9 2 roll ] def
    PFa 0 get PFa 2 get 2 div add PFa 1 get PFa 3 get 2 div add translate
    PFa 2 get -2 div PFa 3 get -2 div PFa 2 get PFa 3 get Rec
    gsave 1 setgray fill grestore clip
    currentlinewidth 0.5 mul setlinewidth
    /PFs PFa 2 get dup mul PFa 3 get dup mul add sqrt def
    0 0 M PFa 5 get rotate PFs -2 div dup translate
	0 1 PFs PFa 4 get div 1 add floor cvi
	{ PFa 4 get mul 0 M 0 PFs V } for
    0 PFa 6 get ne {
	0 1 PFs PFa 4 get div 1 add floor cvi
	{ PFa 4 get mul 0 2 1 roll M PFs 0 V } for
    } if
    stroke grestore } def
/Symbol-Oblique /Symbol findfont [1 0 .167 1 0 0] makefont
dup length dict begin {1 index /FID eq {pop pop} {def} ifelse} forall
currentdict end definefont pop
end
gnudict begin
gsave
0 0 translate
0.100 0.100 scale
0 setgray
newpath
1.000 UL
LTb
350 300 M
63 0 V
3037 0 R
-63 0 V
1.000 UL
LTb
350 507 M
63 0 V
3037 0 R
-63 0 V
1.000 UL
LTb
350 714 M
63 0 V
3037 0 R
-63 0 V
1.000 UL
LTb
350 921 M
63 0 V
3037 0 R
-63 0 V
1.000 UL
LTb
350 1128 M
63 0 V
3037 0 R
-63 0 V
1.000 UL
LTb
350 1335 M
63 0 V
3037 0 R
-63 0 V
1.000 UL
LTb
350 1542 M
63 0 V
3037 0 R
-63 0 V
1.000 UL
LTb
350 1749 M
63 0 V
3037 0 R
-63 0 V
1.000 UL
LTb
350 1956 M
63 0 V
3037 0 R
-63 0 V
1.000 UL
LTb
350 300 M
0 63 V
0 1697 R
0 -63 V
1.000 UL
LTb
715 300 M
0 63 V
0 1697 R
0 -63 V
1.000 UL
LTb
1079 300 M
0 63 V
0 1697 R
0 -63 V
1.000 UL
LTb
1444 300 M
0 63 V
0 1697 R
0 -63 V
1.000 UL
LTb
1809 300 M
0 63 V
0 1697 R
0 -63 V
1.000 UL
LTb
2174 300 M
0 63 V
0 1697 R
0 -63 V
1.000 UL
LTb
2538 300 M
0 63 V
0 1697 R
0 -63 V
1.000 UL
LTb
2903 300 M
0 63 V
0 1697 R
0 -63 V
1.000 UL
LTb
3268 300 M
0 63 V
0 1697 R
0 -63 V
1.000 UL
LTb
1.000 UL
LTb
350 300 M
3100 0 V
0 1760 V
-3100 0 V
350 300 L
LTb
LTb
0.600 UP
0.600 UP
1.000 UL
LT0
LTb
LT0
1613 1847 M
237 0 V
715 507 M
364 207 V
730 414 V
1459 828 V
715 507 Box
1079 714 Box
1809 1128 Box
3268 1956 Box
1731 1847 Box
0.600 UP
1.000 UL
LT1
LTb
LT1
1613 1747 M
237 0 V
715 507 M
364 186 V
365 207 V
365 208 V
365 207 V
364 162 V
365 -27 V
365 304 V
715 507 BoxF
1079 693 BoxF
1444 900 BoxF
1809 1108 BoxF
2174 1315 BoxF
2538 1477 BoxF
2903 1450 BoxF
3268 1754 BoxF
1731 1747 BoxF
0.600 UP
1.000 UL
LT2
LTb
LT2
1613 1647 M
237 0 V
715 507 M
364 174 V
365 121 V
365 68 V
365 -18 V
364 62 V
2903 804 L
365 92 V
715 507 Circle
1079 681 Circle
1444 802 Circle
1809 870 Circle
2174 852 Circle
2538 914 Circle
2903 804 Circle
3268 896 Circle
1731 1647 Circle
1.000 UL
LTb
350 300 M
3100 0 V
0 1760 V
-3100 0 V
350 300 L
0.600 UP
stroke
grestore
end
showpage
}}%
\put(1563,1647){\makebox(0,0)[r]{\small{Rayshooting from files}}}%
\put(1563,1752){\makebox(0,0)[r]{\small{Generate data to files}}}%
\put(1563,1847){\makebox(0,0)[r]{\small{All data in RAM}}}%
\put(1900,50){\makebox(0,0){Processors}}%
\put(100,1180){%
\makebox(0,0)[b]{\shortstack{Speedup}}%
}%
\put(3268,200){\makebox(0,0){ 8}}%
\put(2903,200){\makebox(0,0){ 7}}%
\put(2538,200){\makebox(0,0){ 6}}%
\put(2174,200){\makebox(0,0){ 5}}%
\put(1809,200){\makebox(0,0){ 4}}%
\put(1444,200){\makebox(0,0){ 3}}%
\put(1079,200){\makebox(0,0){ 2}}%
\put(715,200){\makebox(0,0){ 1}}%
\put(350,200){\makebox(0,0){ 0}}%
\put(300,1956){\makebox(0,0)[r]{ 8}}%
\put(300,1749){\makebox(0,0)[r]{ 7}}%
\put(300,1542){\makebox(0,0)[r]{ 6}}%
\put(300,1335){\makebox(0,0)[r]{ 5}}%
\put(300,1128){\makebox(0,0)[r]{ 4}}%
\put(300,921){\makebox(0,0)[r]{ 3}}%
\put(300,714){\makebox(0,0)[r]{ 2}}%
\put(300,507){\makebox(0,0)[r]{ 1}}%
\put(300,300){\makebox(0,0)[r]{ 0}}%
\end{picture}%
\endgroup
\caption{Speedups of new versions of the new microlensing implementation}\label{speedup}
\end{figure}
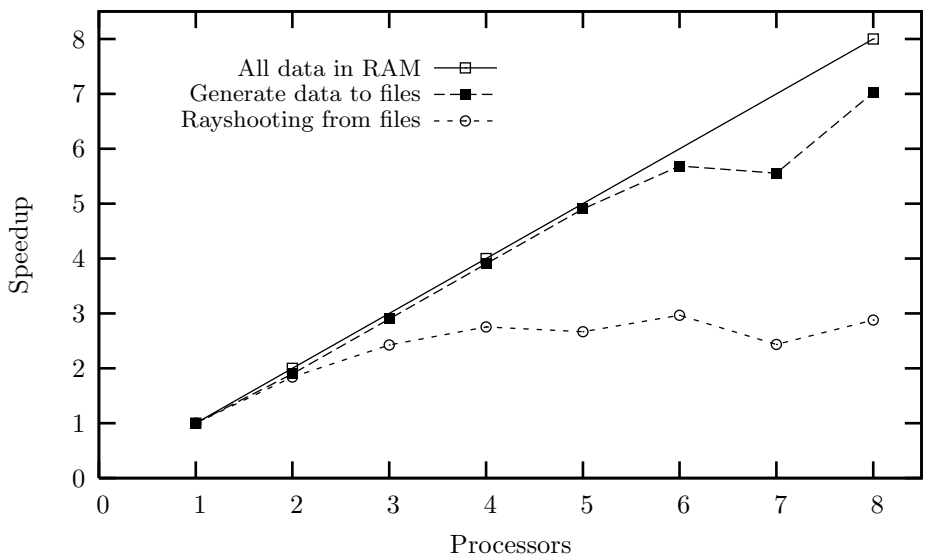

Figure \ref{speedup} shows the speedup obtained for these versions using the new method. 
Version 4 (solid line) uses parallelism during ray-shooting. It achieves a perfect speedup during ray shooting, showing that all the theoretical parallelism that is available can be extracted in practice by the right implementation and in the right circumstances: i.e. each ray is completely independent of the others, there are far many more rays than processes,  our simple ray distribution scheme is achieving a good load balance, and the data is in memory and can be accessed concurrently.

Version 6 uses parallelism in two places: during the generation of stars in the setup phase, and during ray shooting. These are indicated by the dashed and dotted lines respectively. The stars generation shows a reasonable speedup but then tapers off, the shooting phase only gains a speedup of 3. The reason for the poor speedups in this file based implementation is serialization of I/O requests to the data files. To stop this happening it would be necessary to give each process a copy of the files, placed on different disks, and accessed through different I/O controllers. We note again that the  performance of the new approach is now much more dependant on the computer on which it is run. Currently we do not have a different configuration  to test our simulations on, but a 2-3 times speedup is acceptable enough, and in a real situation this particular simulation would use enough RAM so that \emph{all}
the data was cached.

\subsubsection{Summary}

To investigate the properties of quasar microlensing we enhanced a well-known microlensing implementation by massively increasing the number of stars that can be used, and executing some phases of the algorithms in parallel. The simulation is run on supercomputers, data is placed in very large
files, but parallelising the algorithms and adding various caching schemes and a different sort strategy has brought the run-time  to acceptable levels. We can now run simulations in a time comparable to the original method, but using 
10\textsuperscript{9} stars, 100 times more than in the past.

\subsection{Microlensing results}
\begin{figure}[htp]
\centering
\subfigure[2\% = M{$_\odot$}, 98\% = 0.001000\ M{$_\odot$}, total number of stars = 29887581]{
\includegraphics[height=58mm]{map3.eps}}
\subfigure[2\% = M{$_\odot$}, 98\% = 0.000675 M{$_\odot$}, total number of stars = 4920276]{
\includegraphics[height=58mm]{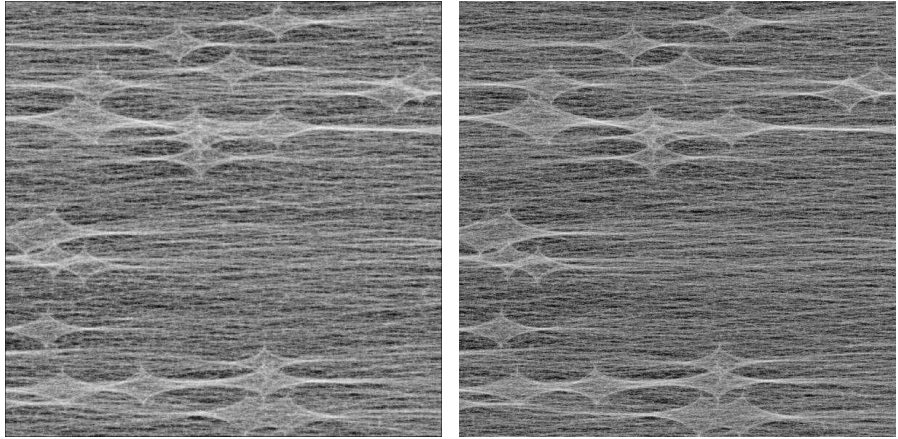}}
\subfigure[2\% = M{$_\odot$}, 98\% = 0.000025 M{$_\odot$}, total number of stars = 1195479475]{
\includegraphics[height=58mm]{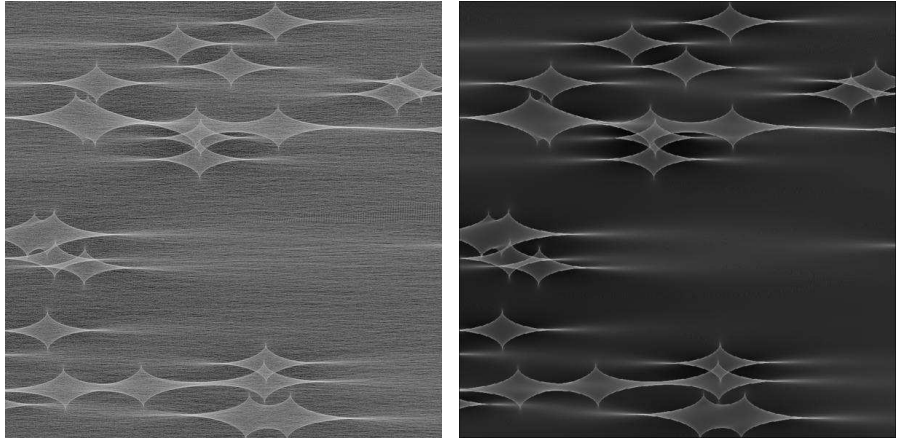}}
\subfigure[2\% = M{$_\odot$}, 98\% = Smooth, total number of stars = 609]{
\includegraphics[height=58mm]{map4.eps}}
\caption{Magnification maps from simulations of bi-modal mass distributions; these use
the same parameters as figure \ref{mag_maps} but use smaller and more numerous masses
  as the compact matter ($\sigma_s$). As the size of the masses is decreased the similarities between
the use of compact masses (a)-(c) and smooth matter (d) become more marked.}\label{mag_maps_big}
\end{figure}
Using our new implementation we can push down the mass size of the compact masses in bi-modal simulations, while at the same time increasing their number. Figure \ref{mag_maps_big} shows simulations of quasar microlensing models that go down to 0.000025 M{$_\odot$} for small compact mass objects; this is about 10 times Earth mass. Future simulations will go to smaller masses and possibly up to 3 billion stars. Of most interest is what happens to the magnification map as the masses become smaller. Figure \ref{mag_maps_big} begins at case (c) of figure \ref{mag_maps} and reduces the mass to 0.000025 M{$_\odot$}, increasing the number of stars to over a billion. With these small masses the large-scale caustics are becoming clearer and the fine-structure in-between the caustics is becoming smoother. Initial analysis indicates that as the small masses become smaller and more numerous the  magnification distribution will tend to that of the smooth-matter case (d), but the size of the source will be important.
Results of more complete investigations will be reported in forthcoming papers.

\section{Discussion}
Our new approach is of interest in itself, and can be studied as an instance of a large, parallel, computationally intensive implementation. Although our current aims excluded modifying the physics code, that can still be done at some later time.

\subsection{Software Engineering}
The following could be worthwhile scientific research projects based around our implementation:

{\bf Optimum distribution of rays to processors.} Do rays that are close to clumps of matter really take longer to process than those that aren't? And by how much? How close to optimal
is our simple static load-distribution scheme?

{\bf Dynamic load balancing}. Instead of fixing the allocation of rays to processors, the allocation could be varied at run-time, so that rays are moved from heavily loaded processors to lightly loaded processors. We could implement microtasking where each ray could be a task, with a scheduler pulling jobs off a run-queue and assigning them to processors for execution. We have not investigated whether the load on processors varies  enough to warrant the overhead of dynamic load-balancing; our impressions are that it does not, for very large runs.

{\bf Distributed memory.} A simulation of billions of stars generates hundreds of gigabytes of data, and this data has to go somewhere. If there are enough processes available with enough total memory, the data can be distributed amongst them and fetched across a network.

{\bf “Pre-set” stars.} It would useful to generate stars and cells that can be saved and re-used, thus skipping the initialization phase altogether. A standard for the storing of lens/cell data could be developed by the lensing community,  so that stars and cells can be stored and then simulated, modified, and re-used as necessary. 

{\bf Multithreaded version.} The new implementation  has been built with this in mind. There is little  data  specific to a process that would have to be made specific to a thread; the significant shared data is external and can be accessed just the same by multiple threads or processes. Therefore a multithreaded version could be generated.

{\bf Use a commercial database for cells and stars.} The assumption is that a commercial database will be more efficient at general file caching and perhaps can be tuned to the usage patterns of our simulations. 

{\bf File compression.} Compressing the stars and cell files on the fly will not make a simulation run faster, in fact it will run slower, but the data compresses to about 63\% of its original size. This would be useful if we run short of disk space or go to bigger simulations, but particularly it could be useful for a distributed-memory version.

\subsection{Science}
Beyond  computer engineering,  there are the scientific aspects of simulating microlensing on computers, for example with the use of a cell tree. The cell tree is one reason why the original method executed so speedily, but it is also the reason for the huge amount of data generated. Instead of using a tree it may be possible to  use other more recent models such as adaptive meshes that are used in N-body simulations \citep{yahagi}. 

There are also enhancements that can be made that add more physics to the implementation, such as implementing moving sources and moving stars. It is clear that both the source and the lens and the objects that make up the lens, if it is complex, are moving; and the gravitational potential and light curves are changing over time. The ability to simulate time-changing lensing events would be a significant advance over all current approaches.

\section{Conclusion}

We have generated a new version of a ray-shooting microlensing simulation tool  that can be used to run huge simulations and continue  research in quasar dark matter microlensing. We have discovered that the ray-shooting is easily
parallelizable and simple load-balancing works well; parallelizing the
stars and cell generation  is not so easy, and memory caching could
be improved with more research into the patterns of how the data is accessed.

Our new approach will now be used to continue research in microlensing; based on its performance and usefulness we can determine what modifications to pursue for the future.

\section{Acknowledgements}

We thank the anonymous reviewer for comments that improved the quality of
this paper.

\end{document}